\documentclass[prd,superscriptaddress,floatfix,nofootinbib,twocolumn]{revtex4-2}
\pdfoutput=1 
\usepackage[T1]{fontenc} 
\usepackage{amsmath,amssymb,amsfonts,amsthm}
\usepackage{graphicx}
\usepackage[usenames,dvipsnames]{color}
\usepackage{enumitem}
\usepackage{verbatim}
\usepackage[percent]{overpic}
\usepackage{rotating}
\usepackage[colorlinks=true]{hyperref}
\usepackage{array}
\usepackage{mathtools}
\usepackage{lmodern}
\usepackage[normalem]{ulem}
\usepackage{braket}
\usepackage{tensor}
\usepackage{dsfont}
\usepackage{bm}
\usepackage{tikz}

\usepackage{mathrsfs}

\usetikzlibrary{decorations.pathmorphing}
\usepackage{CJKutf8}

\DeclareMathOperator\arctanh{arctanh}

\definecolor{patriarch}{rgb}{0.5, 0.0, 0.5}
\definecolor{darkraspberry}{rgb}{0.53, 0.15, 0.34}
\definecolor{brinkpink}{rgb}{0.98, 0.38, 0.5}
\definecolor{skobeloff}{rgb}{0.0, 0.48, 0.45}
\definecolor{mypink}{RGB}{226,68,130}
\definecolor{darkpastelgreen}{rgb}{0.01, 0.75, 0.24}
\definecolor{pigmentgreen}{rgb}{0.0, 0.65, 0.31}

\newcommand{\mr}[1]{{\color{mypink}#1}}

\usetikzlibrary{positioning}
\usetikzlibrary{calc,through,backgrounds}

\newcommand{\ii}{\mathsf{i}}

\begin{abstract}
    Previous studies have shown that an Unruh-DeWitt (UDW) detector, when coupled linearly to a massless scalar field and permitted to fall radially into certain black holes, will exhibit nonmonotonicity in its transition properties near the horizon. Specifically, the transition probability of a detector freely falling into a (3$+$1)-dimensional Schawrzschild black hole, when considering the Unruh and Hartle-Hawking vacuum states, was shown to possess a local extremum at horizon crossing [K. K. Ng \textit{et al.}, New J. Phys. 24, 103018 (2022)]. The transition rate of a detector falling into a static (2$+$1)-dimensional Bañados-Teitelboim-Zanelli (BTZ) black hole, for the Hartle-Hawking state, was also found to have multiple local extrema near the horizon under certain parameter settings [M.R. Preciado-Rivas \textit{et al.}, Phys. Rev. D 110, 025002 (2024)]. These discoveries are of interest, as they suggest that the event horizon of a black hole may be discerned by a local probe when quantum field theory effects are included. In this paper, we explore the problem of a UDW detector falling freely into a rotating BTZ black hole. We numerically compute the detector’s transition rate for different values of black hole mass, black hole angular momentum, detector energy gap, and field boundary conditions at infinity. Our results lead to a more generalized description of the behavior of particle detectors in BTZ black hole spacetime, from which the previous nonrotating BTZ case can be retrieved in the limit as angular momentum vanishes.
\end{abstract}

\begin{document} 

\title{Singular excitement beyond the horizon of a rotating black hole}

\author{Sijia Wang}
\email{s676wang@uwaterloo.ca}
\affiliation{Department of Physics and Astronomy, University of Waterloo, Waterloo, Ontario, N2L 3G1, Canada}

\author{María R. Preciado-Rivas}
\email{mrpreciadorivas@uwaterloo.ca} 
\affiliation{Department of Applied Mathematics, University of Waterloo, Waterloo, Ontario, N2L 3G1, Canada}
\affiliation{Institute for Quantum Computing, University of Waterloo, Waterloo, Ontario, N2L 3G1, Canada}

\author{Massimiliano Spadafora}
\email{massi.spadafora@gmail.com}
\affiliation{Department of Physics, University of Calabria, Rende, Calabria, 87036, Italy}

\author{Robert B. Mann}
\email{rbmann@uwaterloo.ca}
\affiliation{Department of Physics and Astronomy, University of Waterloo, Waterloo, Ontario, N2L 3G1, Canada}
\affiliation{Institute for Quantum Computing, University of Waterloo, Waterloo, Ontario, N2L 3G1, Canada}
\affiliation{Perimeter Institute for Theoretical Physics,  Waterloo, Ontario, N2L 2Y5, Canada}



\maketitle
\flushbottom


\section{Introduction}\label{sec:introduction}
In the absence of a quantum theory of gravity, relativistic quantum information provides valuable technical and conceptual insight for investigating quantum effects in curved spacetime.
  One common operational approach in relativistic quantum information is to use an idealized particle detector to probe quantum fields, particularly in situations where the notion of a particle is ill defined. The simplest such detector is the Unruh-DeWitt (UDW) detector~\cite{Unruh.effect, DeWitt1979}, a two-level quantum system that couples locally to a quantum scalar field. One quantity of interest in this framework is the transition probability between the two levels of the detector; another is the derivative of this probability with respect to total detection time, known as the transition rate. Some well-known results obtained using the UDW detector model are the Unruh effect~\cite{Unruh.effect,DeWitt1979,Davies1975}, wherein an accelerated detector in flat spacetime exhibits a thermal response proportional to its acceleration, and entanglement degradation~\cite{Fuentes-Schuller:2004iaz}, which occurs when two detectors are in relative noninertial motion.

The particle detector model has also been used to study black hole spacetimes. Some results obtained in this vein are described in~\cite{Israel1976}, for the static detector outside of a Schwarzschild black hole, which was found to respond thermally to the Hartle-Hawking-Israel state~\cite{HartleHawking1976}; and in~\cite{hodgkinsonStaticStationaryInertial2012}, for a stationary detector corotating with a Bañados-Teitelboim-Zanelli (BTZ) black hole, which was found to thermalize at the Hawking temperature.

Much less is known about the response of detectors that freely fall into a black hole and cross its horizon. The general expectation is that the transition rate will be smooth and the detector will not thermalize. 
It has been shown that  a detector freely falling 
across the horizon of a Bertotti-Robinson spacetime 
has a  transition rate that linearly decreases as the horizon is crossed \cite{Conroy_2022}.
Several studies have been carried out thus far in lower-dimensional settings. References \cite{Juarez-Aubry:2014jba,JuarezPhDThesis} numerically compute the transition rate of a UDW detector freely falling into a (1+1)-dimensional Schwarzschild black hole, finding evidence that the thermal properties of the detector are gradually lost during infall. In~\cite{Juarez-Aubry:2021tae}, the transition rate of a detector infalling toward a (1+1)-dimensional Cauchy horizon was shown to diverge for a field in both the Unruh and Hartle-Hawking-Israel vacuum states. Another study~\cite{Gallock-Yoshimura:2021yok} has shown that it is possible for two infalling detectors to harvest entanglement in (1+1)-dimensional Schwarzschild spacetime, even when causally separated by the black hole horizon. 
The response rate of a detector falling radially in a (2+1)-dimensional static BTZ spacetime was computed in~\cite{hodgkinsonStaticStationaryInertial2012}, but the detector had only been switched on outside of the horizon.

Recently, the response function of a detector initially at rest at infinity 
and falling freely toward and across the horizon of a (3+1)-dimensional Schwarzschild black hole was computed~\cite{Ng2022}
for a scalar field in the Unruh and Hartle-Hawking-Israel states. 
In both cases, contrary to expectation, a local extremum was observed in the transition rate as the detector crossed the horizon. The locus of the extremum was found to depend on the interaction duration and the detector's energy gap. This result is of significant interest, as it suggests that a local probe knows when it is crossing the Schwarzschild horizon.

The transition rate of a freely falling detector in (2+1)-dimensional nonrotating BTZ spacetime was computed in~\cite{MariaRosaBTZ}, exhibiting similar, but richer, behavior compared to the Schwarzschild case. In particular, the transition rate as a function of detector proper time was shown to oscillate slowly outside the black hole, but fluctuate more dramatically near the horizon for certain ranges of the black hole's mass and the detector's initial position. The transition rate then increases and diverges at the singularity. 
Moreover, the transition rate was found to be nondifferentiable at certain detector proper times; these cusps in the transition rate were dubbed ``glitches.'' Theoretically, one could use these glitches as an early warning system as to whether or not the observer is about to fall into the horizon. Both the profile of the detector transition rate and the locations of the glitches are highly dependent on the black hole mass and the detector's initial position.

Expanding on the static BTZ black hole research, we investigate the transition rate of a UDW detector falling into a rotating BTZ black hole in (2+1) dimensions. The rationale for this investigation is threefold:

\begin{enumerate}
    \item We wish to understand the effect of frame dragging on the response of an infalling detector.
    \item We anticipate the possibility of new behaviors in detector transition rate when rotation is considered, particularly when the black hole approaches extremality. This intuition is informed by previous research showing that entanglement harvesting can be amplified by black hole rotation~\cite{Robbins2022}, especially as angular momentum increases to the extremal limit.
    \item The detector transition rate for a BTZ black hole is efficient to compute. Rather than the mode sum approach used in the (3+1)-dimensional Schwarzschild case~\cite{Ng2022}, which is computationally intensive, we can calculate the transition rate as a much simpler image sum.
\end{enumerate}

We find that the detector transition rate curves for the rotating BTZ spacetime take on a similar shape as in the nonrotating BTZ case, but the rotation of the black hole pushes the features in the curves further away from the singularity. However, the transition rate of the detector \textit{does not} diverge at the time the detector reaches its point of closest approach to the
black hole's singularity when rotation is present. Additionally, new glitch behaviors arise in the rotating case. In the nonrotating BTZ case, the detector encounters only one set of glitches, which are symmetric in the image sum. ``Symmetric'' in this context means that, if a glitch occurs at a proper time $\tau_g$ in term $n$ of the image sum, it will also occur at the same time $\tau_g$ in term $-n$. In the rotating BTZ spacetime, by contrast, the detector encounters four types of glitches---two of which are symmetric in $n$ and two of which are asymmetric. 
The sets of glitches present different behaviors for small $|n|$, but all asymptote monotonically toward the time to the inner horizon  (or a reflection of this about  $\Delta\tau/\ell=\pi/2$) as $n\rightarrow \pm\infty$. One can see how the special case of the nonrotating BTZ spacetime may be recovered from the rotating BTZ setting: in the limit of vanishing black hole angular momentum,  the time to the inner horizon and its reflection go to the singularity; thus, the glitches are collapsed into the singularity as $n\to\pm\infty$.
The pattern of the glitches in the transition rate provides us with a predictor for the structure of the spacetime  since it depends on the black hole mass and the detector's switch-on moment, allowing us to anticipate if the detector will encounter the black hole horizons.

The outline of our paper is as follows. In \autoref{sec:setup}, we present the basic formalism needed to compute the transition rate of a UDW detector freely falling in a rotating BTZ spacetime. We present in \autoref{sec:results} the results of our study, showing how the transition rate depends on the mass of the black hole, the angular momentum of the black hole, the boundary conditions of the field at asymptotic infinity, and the energy gap of the detector. We also present the locations of the four types of glitches in the rotating case, as well as the transition rate behavior at the time the detector reaches its closest approach to the black hole singularity. In \autoref{sec:conclusion}, we draw conclusions from the results and propose directions for future research.

\section{Setup}\label{sec:setup}

We consider the Unruh-DeWitt model for a detector coupled to a quantum massless scalar field. Since much of the formalism has already been derived~\cite{hodgkinsonStaticStationaryInertial2012}, we shall only reiterate the ideas relevant to our study.

The UDW detector that we consider is a pointlike qubit, having two states denoted by $|0\rangle$ and $|E\rangle$. The energy eigenvalues of the states are $0$ and $E$, respectively, where $E$ may be positive or negative. The detector starts out in the $\ket{0}$ state, and a positive (negative) $E$ indicates that $\ket{0}$ is the detector's ground (excited) state. The detector moves on a timelike worldline $\mathsf{x}(\tau)$, parametrized by its proper time $\tau$, and interacts with a massless scalar field $\hat{\phi}$ via the interaction Hamiltonian
\begin{equation}
\hat{H}_{\mathrm{int}}=\lambda\chi(\tau)\hat{\mu}(\tau)\otimes\hat{\phi}(\tau)\,,
\end{equation}
where $\hat{\mu}(\tau)=\ket{E}\bra{0}e^{\ii E\tau}+\ket{0}\bra{E}e^{-\ii E\tau}$ is the detector's monopole moment operator, $\lambda$ is a coupling constant, and $\chi$ is the switching function, specifying how the detector is switched on and off.

Perturbatively, the probability that the detector makes a transition from the state $\ket{0}$ to the state $\ket{E}$ is proportional to the response function, given by~\cite{Birrell_Davies_QFTCS}
\begin{equation}
\mathcal{F}=\int\mathrm{d}\tau'\mathrm{d}\tau''\mathrm{e}^{-\mathrm{i}E(\tau'-\tau'')}W(\tau',\tau''),
\end{equation}
to leading order, 
where $W$ is the pullback along the detector's worldline of the field's Wightman function in the state in which the field was initially prepared~\cite{kay_theorems_1991,Fewster:1999gj,Junker:2001gx,Louko:2007mu}. In (2+1) spacetime dimensions, we may take $\chi$ to be the characteristic function of a time interval (i.e., sharp switching) since $\mathcal{F}$ remains well defined despite the nonsmoothness at the switch-on and switch-off moments ~\cite{hodgkinsonStaticStationaryInertial2012}. Moreover, we may consider the detector's transition rate, given by (a multiple of) the derivative of $\mathcal{F}$ with respect to the switch-off moment. The transition rate in the sharp switching limit and to leading order 
in the coupling constant is~\cite{hodgkinsonStaticStationaryInertial2012}
\begin{equation}
\dot{\mathcal{F}}_\tau=\frac{1}{4}+2 \int_0^{\Delta \tau} \mathrm{d} s \operatorname{Re}\left[\mathrm{e}^{-\mathrm{i} E s} W(\tau, \tau-s)\right], \label{E2}
\end{equation}
where $\tau$ denotes the switch-off moment and $\Delta \tau$ is the total proper time that the detector operates. $\Delta \tau=\tau-\tau_0$, where $\tau_0$ is the switch-on moment. The transition rate can be interpreted in terms of an ensemble of identical detectors, all following the trajectory $\mathsf{x}(\tau)$; it tells us the fraction of detectors per unit time that have undergone a transition when observed at time $\tau$. To measure $\dot{\mathcal{F}}$ as a function of $\tau$ requires a set of identical ensembles, each observed at different values of $\tau$~\cite{Louko:2007mu,Satz2007}.

The line element of the rotating BTZ spacetime is
\begin{equation}
    ds^2 = - h(r) dt^2  + \frac{1}{h(r)} dr^2 + r^2 \left[ N_\varphi (r) dt + d\varphi \right]^2,\label{E3}
\end{equation}
where
\begin{align}
    h(r) &= -M + \frac{r^2}{\ell^2} + \frac{J^2}{4r^2},\\
    N_\varphi (r) &= - \frac{J}{2r^2}.
\end{align}
$M$ is the (dimensionless) mass of the black hole, and $J$ is its angular momentum. Additionally, $t\in\mathbb{R}$, $r\in(0,\infty)$, and $\varphi\in[0,2\pi)$. This metric (\ref{E3}) is a vacuum solution to the Einstein field equations with cosmological constant $\Lambda=-1/\ell^2$, where $\ell~(>0)$ is the anti-de Sitter (AdS) length. The outer and inner horizons (where $h(r)=0$) are given by the expressions
\begin{equation}
    r^2_{\pm} = \frac{1}{2}\left( M \ell^2 \pm \sqrt{M^2 \ell^4 - \ell^2 J^2}\right).\label{E6}
\end{equation}

By setting $M=1$ (without loss of generality) and $\varphi\rightarrow y\in(-\infty,\infty)$ in (\ref{E3}), one obtains the AdS$_3$-Rindler spacetime. Conversely, the BTZ spacetime can be obtained from AdS$_3$-Rindler by identifying $y\rightarrow\phi\in(0,\sqrt{M})$, and then rescaling $\phi$ and the radial and time coordinates to obtain (\ref{E3}); it is therefore locally equivalent to anti-de Sitter spacetime. 
We consider this fact when dealing with a quantum field in BTZ spacetime, where the correlation functions can be represented as a sum of correlators in AdS$_3$.

 We pause to comment on the nature of the BTZ spacetime at $r=0$. 
   In both the rotating and nonrotating cases, the curvature of the BTZ spacetime is constant, with a well-defined Kretschmann scalar at every point, including as $r\to 0$. However the identification that produces the BTZ spacetime is performed along a spacelike Killing vector, which is simply $\partial_\phi$ for the static BTZ black hole and a function of $\partial_\phi$ and $\partial_t$ for the rotating BTZ black hole.

These Killing vectors can become null and then timelike in some regions, which  must be excised  from the covering anti-de Sitter space to make the identifications valid; otherwise, the spacetime will have closed timelike curves \cite{BTZ2}. This, in turn, has the consequence that 
  the spacetime becomes geodesically incomplete and the surface $r=0$ becomes a singularity in the causal structure \cite{BTZ2,carlipBTZ}.
If $J=0$, the manifold   is additionally non-Hausdorff at $r=0$ \cite{BTZ2}.
If the black hole is formed from collapsing matter then a genuine curvature singularity appears at $r=0$ \cite{Ross:1992ba}.

We consider a freely falling detector that is initially corotating with the black hole. The timelike geodesic followed by the detector is given in BTZ coordinates by
\begin{align} \label{eq:trajectory}
    r(\tau) &= r_0 \sqrt{1 - \left[ 1 - \left(\frac{r_- r_+}{r_0^2}\right)^2 \right] \sin^2{\left( \frac{\tau}{\ell}  \right)}}, \nonumber\\
    \begin{split}
        t(\tau) &= \frac{\ell^2}{r_+^2-r_-^2} \left( r_+ \arctanh{ \left[ \frac{r_+}{r_0}\sqrt{\frac{r_0^2-r_-^2}{r_0^2-r_+^2}}\tan{\left( \frac{\tau}{\ell}\right)}\right]} \right.\\
        & \ - r_- \left. \arctanh{ \left[ \frac{r_-}{r_0}\sqrt{\frac{r_0^2-r_+^2}{r_0^2-r_-^2}}\tan{\left( \frac{\tau}{\ell}\right)}\right]} \right),
    \end{split}\\
    \begin{split}
        \varphi(\tau) &= \frac{\ell}{r_+^2-r_-^2} \left( r_- \arctanh{ \left[ \frac{r_+}{r_0}\sqrt{\frac{r_0^2-r_-^2}{r_0^2-r_+^2}}\tan{\left( \frac{\tau}{\ell}\right)}\right]} \right. \\
        & \ - r_+ \left. \arctanh{ \left[ \frac{r_-}{r_0}\sqrt{\frac{r_0^2-r_+^2}{r_0^2-r_-^2}}\tan{\left( \frac{\tau}{\ell}\right)}\right]}\right),
    \end{split} \nonumber
\end{align}
where $r_0$ is the radial position at which the detector is switched on, and constant terms in the temporal and angular coordinates have been omitted. To simplify the notation, we will write $\tau/\ell$ as $\tau$.

We consider a massless conformally coupled quantum scalar field $\hat\phi(\mathsf{x})$ satisfying the Klein-Gordon equation,
\begin{equation}
    (\Box-R/8)\hat\phi(\mathsf{x})=0\,,
\end{equation}
where $\Box$ is the d'Alembert operator, and $R$ is the Ricci scalar. The Wightman function $W(\mathsf{x},\mathsf{x}'):=\langle 0|\hat\phi(\mathsf{x})\hat\phi(\mathsf{x}')|0 \rangle$ is the two-point correlation function in the vacuum state $|0\rangle$ of the field. Since AdS$_3$ spacetime is  locally equivalent to BTZ spacetime, we may express the field correlations in BTZ as an image sum of the Wightman function in AdS$_3$ spacetime. The specific form is~\cite{LifschytzBTZ,carlipBTZ}
\begin{equation}
    \begin{split}
        & W_{\text{BTZ}}(\mathsf{x},\mathsf{x}') = \sum_{n=-\infty}^\infty W_{\text{AdS}}(\mathsf{x},\Gamma^n\mathsf{x}') \\
        & =\frac{1}{4\pi\sqrt{2}\ell} \sum_{n=-\infty}^\infty \left[\frac{1}{\sqrt{\sigma_\epsilon(\mathsf{x},\Gamma^n\mathsf{x}')}}-\frac{\zeta}{\sqrt{\sigma_\epsilon(\mathsf{x},\Gamma^n\mathsf{x}')+2}}\right],
    \end{split}
\end{equation}
where $\Gamma:(t,r,\varphi)\mapsto(t,r,\varphi+2\pi)$. The parameter $\zeta\in\{-1,0,1\}$ specifies the boundary conditions of the field at asymptotic infinity: Neumann, transparent, or Dirichlet, respectively. The Wightman function with Neumann or Dirichlet boundary conditions corresponds to a Kubo-Martin-Schwinger (KMS) state 
and is analytic outside the black hole horizon, thus characterizing a Hartle-Hawking state~\cite{Avis:1978,LifschytzBTZ}. On the other hand, the state with the transparent boundary condition does not have a clear physical meaning~\cite{Avis:1978,LifschytzBTZ}.  

$\sigma_\epsilon(\mathsf{x},\mathsf{x}')$ is the squared geodesic separation (scaled by $\ell^2$) between two points in the covering AdS$_3$ spacetime, and the $\epsilon$ notation encodes the distributional character of the Wightman function in the limit as $\epsilon\rightarrow 0_+$. Let the unprimed coordinates correspond to $\mathsf{x}(\tau)$ on the detector trajectory \eqref{eq:trajectory} and the primed coordinates to $\mathsf{x}(\tau-s)$. The explicit form of $\sigma_\epsilon(\mathsf{x}(\tau),\Gamma^n\mathsf{x}(\tau-s))$ is then
\begin{equation}
    \begin{split}
        \sigma_\epsilon(\mathsf{x}(\tau)&,\Gamma^n\mathsf{x}(\tau-s)) = \\ 
        & -1 + \cos{(\tau - s)} \left[ \cos{(\tau)} K_{n,1} + \sin{(\tau)} K_{n,2}\right] \\
        & + \sin{(\tau-s)}  \left[ \sin{(\tau)} K_{n,3} + \cos{(\tau)} K_{n,4}  \right],
    \end{split}
    \label{eq:geodesic_dist}
\end{equation}
which we will denote in short as $\sigma_{\epsilon,n}(\tau,\tau-s)$, where
\begin{widetext}
\begin{align}
    & K_{n,1}=\frac{1}{r_+^2 - r_-^2} \left[ \cosh{\left(2 \pi n \frac{r_+}{\ell} \right)} \left(r_0^2-r_-^2\right) - \cosh{\left(2 \pi n \frac{r_-}{\ell} \right)} \left(r_0^2-r_+^2\right) \right] \\
    & K_{n,2} = \frac{1}{r_0} \frac{\sqrt{\left(r_0^2-r_+^2 \right) \left( r_0^2-r_-^2\right)}}{r_+^2 - r_-^2} \left[ \sinh{\left(2 \pi n \frac{r_+}{\ell} \right)} r_- - \sinh{\left(2 \pi n \frac{r_-}{\ell} \right)} r_+  \right] \\
    & K_{n,3} = \frac{1}{r_0^2} \frac{1}{r_+^2 - r_-^2} \left[ \cosh{\left(2 \pi n \frac{r_-}{\ell} \right)} r_+^2 \left( r_0^2  - r_-^2 \right) - \cosh{\left(2 \pi n \frac{r_+}{\ell} \right)}  r_-^2 \left( r_0^2  - r_+^2 \right) \right] \\
    & K_{n,4} =\frac{1}{r_0} \frac{\sqrt{\left(r_0^2-r_+^2 \right) \left( r_0^2-r_-^2\right)}}{r_+^2 - r_-^2} \left[ \sinh{\left(2 \pi n \frac{r_-}{\ell} \right)} r_+  -  \sinh{\left(2 \pi n \frac{r_+}{\ell} \right)} r_- \right] = - K_{n,2}.
    \label{eq:kns}
\end{align}
The transition rate is then
\begin{equation}
    \dot{\mathcal{F}}_\tau=\frac{1}{4}+\frac{1}{2\pi\sqrt{2}} \sum_{n=-\infty}^\infty \int_0^{\Delta\tau} \mathrm{d} s \operatorname{Re}\left[\mathrm{e}^{-\mathrm{i}Es}\left(\frac{1}{\sqrt{\sigma_{\epsilon,n}(\tau,\tau-s)}}-\frac{\zeta}{\sqrt{\sigma_{\epsilon,n}(\tau,\tau-s)+2}}\right)\right]
    .
    \label{eq:transition_rate}
\end{equation}
\end{widetext} 

Several observations can be made about the rotating case relative to the nonrotating case described in~\cite{MariaRosaBTZ}.

First, from (\ref{eq:transition_rate}) when calculating $\dot{\mathcal{F}}_\tau$, the $n\neq 0$ terms are not invariant under $n\rightarrow -n$. This is different from the nonrotating case, where there was invariance in $n\rightarrow -n$. Thus, the $n$ and $-n$ terms must be summed separately in the rotating case.

Second, while the transition rate formula (\ref{eq:transition_rate}) is derived using expression (\ref{eq:geodesic_dist}), which is valid only in the black hole exterior, (\ref{eq:transition_rate}) holds over the detector’s full trajectory, even when the detector operates after entering the black hole. This follows from the existence of a global analytic chart, by analytic continuation in (\ref{eq:transition_rate}). This analytic continuation can be performed in both the rotating and nonrotating cases.

As in the nonrotating case, the transition rate decomposes as
\begin{equation}
\dot{\mathcal{F}}_\tau=\dot{\mathcal{F}}_\tau^{n=0}+\dot{\mathcal{F}}_\tau^{n\neq 0},
\end{equation}
where $\dot{\mathcal{F}}_\tau^{n=0}$ consists of the $n=0$ term, and $\dot{\mathcal{F}}_\tau^{n\neq 0}$ consists of the sum over the $n\neq 0$ terms. $\dot{\mathcal{F}}_\tau^{n=0}$ is the transition rate of the detector in AdS$_3$. 

The square roots in (\ref{eq:transition_rate}) are positive for positive arguments, and they are analytically continued to negative arguments by giving $s$ a small negative imaginary part~\cite{hodgkinsonStaticStationaryInertial2012,kay_theorems_1991}. This procedure is valid in both the rotating and nonrotating cases.

For certain values of $s$, the right-hand side of (\ref{eq:geodesic_dist}) vanishes. Specifically, $\sigma_{\epsilon,n}(\tau,\tau-s)$ vanishes for
\begin{equation}
    \begin{split}
    s_{1,2}^*(\tau) = \tau
     - \operatorname{arctan2} & \left(  {a \mp b \sqrt{a^2 + b^2 - 1}}, \right. \\
       & \hspace{0.58em} \left.{b \pm a \sqrt{a^2 + b^2 - 1}}\right) ,
    \end{split}
\end{equation}
where $a=\sin{(\tau)} K_{n,3} + \cos{(\tau)} K_{n,4}$, $b=\cos{(\tau)} K_{n,1} + \sin{(\tau)} K_{n,2}$, and $\operatorname{arctan2}(y,x)$ is the argument of $x+\ii y$. Similarly, the quantity $\sigma_{\epsilon,n}(\tau,\tau-s)+2$ in the second term of the integrand vanishes for
\begin{equation}
    \begin{split}
    s_{3,4}^*(\tau) = \tau - \operatorname{arctan2}&\left( {-a \mp b \sqrt{a^2 + b^2 - 1}},\right. \\
    & \hspace{0.41em} \left. {-b \pm a \sqrt{a^2 + b^2 - 1}}\right) .
    \end{split}
\end{equation}
These values of $s$ yield singularities in the integrand of \eqref{eq:transition_rate}. However, since they are of the inverse square root type, they are still integrable. 

When $s_{1,2,3,4}^*(\tau)$ is equal to one of the bounds of  integration in (\ref{eq:transition_rate}) (i.e., $s_{\mr{1,3}}^*(\tau)=\mr{\tau}$ or $s_{\mr{2,4}}^*(\tau)=\mr{0}$), the transition rate function will gain a jump discontinuity in its first derivative at time $\tau$. 
Such nondifferentiable points in the transition rate, dubbed ``glitches,'' were previously observed in the case of a UDW detector falling into a nonrotating BTZ black hole~\cite{MariaRosaBTZ}.

For the static black hole, the glitches encountered by the detector as it falls toward the singularity originate exclusively from the $s_1^*(\tau)=\tau$ case---that is, when $\sigma_{\epsilon,n}(\tau,\tau-s)$ vanishes at the upper bound $\tau$. The singularities $s^*_1(\tau)$ in the integrand have a geometric origin, occurring when the detector's trajectory crosses the future light cone of the switch-on event. There are many discrete values of the proper time at which this can happen because the light cones wrap around the $\varphi$ direction. 

For the rotating black hole,  
all four types of singularities, $s_1^*(\tau)$, $s_2^*(\tau)$, $s_3^*(\tau)$, and $s_4^*(\tau)$, in the integrand contribute in the domain over which the detector is switched on. We see that we then have three new types of glitches.
The explicit locations of all glitches are
respectively
\begin{align}
    & \tau_{1}^{*}  = \arctan{\left( \frac{K_{n,2} + K_{n,1} \sqrt{K_{n,2}^2 + K_{n,1}^2 - 1}}{K_{n,1} - K_{n,2} \sqrt{K_{n,2}^2 + K_{n,1}^2 - 1}}\right) } \label{eq:glitch-explicit-tau1}, \\
    & \tau_2^* = \arcsin{\left( \sqrt{\frac{1 - K_{n,1}}{K_{n,3}-K_{n,1}}} \right)}, \\
    & \tau_{3}^{*}  = \arctan{\left( \frac{-K_{n,2} + K_{n,1} \sqrt{K_{n,2}^2 + K_{n,1}^2 - 1}}{-K_{n,1} - K_{n,2} \sqrt{K_{n,2}^2 + K_{n,1}^2 - 1}}\right) }, \\
    & \tau_4^* = \arcsin{\left( \sqrt{\frac{1 + K_{n,1}}{K_{n,1}-K_{n,3}}} \right)},
    \label{eq:glitch-explicit-tau4}
\end{align}
where the glitches at $\tau_{1}^{*}$ have the same origin as in the static case,  $s^*_1$ being coincident with the upper bound of integration (i.e., $s^*_1(\tau)=\tau$). The glitches at $\tau_{3}^{*}$ arise from $s^*_3$
being coincident with the upper bound of integration, which is $s_3^*(\tau)=\tau$, but now correspond to the detector’s trajectory crossing a future-directed timelike curve beginning at the switch-on event. This is a new feature absent in the static case. The glitches at $\tau_{2}^{*}$ and $\tau_{4}^{*}$ arise from $s_{2}^{*}$ and $s_{4}^{*}$ being  coincident with  the lower bound of integration (i.e., $s_{2,4}^*(\tau)=0$)---another feature that is absent in the static case. 
Those from $\tau_{2}^{*}$ occur when the detector's trajectory crosses the future light cone of the switch-on event, whereas those from $\tau_{4}^{*}$ occur when the detector intersects a future-directed timelike curve beginning at the switch-on event. 
For all four types of glitches, both the light cones and the future-directed timelike curves can wrap around the $\varphi$ direction, leading to many possible discrete glitches.

\section{Results}\label{sec:results}

In this section, we present our results for the transition rate of the detector with different black hole angular momenta, boundary conditions of the field, masses of the black hole, and energy gaps of the detector. We also examine the glitches in the rotating case versus the nonrotating case. 
In all our results, the initial position of the detector is $r_0 = 100\,r_+$, and we have chosen the detector to be switched on when it starts to freely fall, that is, $\tau_0/\ell=0$. Hence, the switch-off time or the total detection time coincides with the proper time of the detector, $\Delta\tau/\ell = \tau/\ell$. Consequently, the detector is at its initial position $r=r_0$ when $\Delta\tau/\ell=0$, and it reaches $r=\frac{r_{-}r_{+}}{r_0} = 10^{-2} r_{-}$, 
its distance of closest approach to the  
(geodesic incompleteness) singularity when $\Delta\tau/\ell=\pi/2$.


In all of our computations, we truncated the image sums at a term whose absolute value over the entire detection time interval was on the order of one part in 10000 of the cumulative total, even in the worst-case scenario. We only required 1000 terms inside the outer horizon and 200 terms outside of it in any image sum to obtain curves that did not visually change upon adding more terms.

\subsection{Transition rate for different angular momenta and boundary conditions}

\begin{figure}
    \centering
    \includegraphics[width=\linewidth]{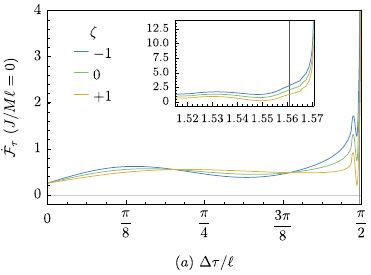}
    \includegraphics[width=\linewidth]{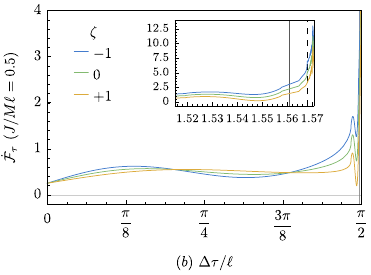}
    \includegraphics[width=\linewidth]{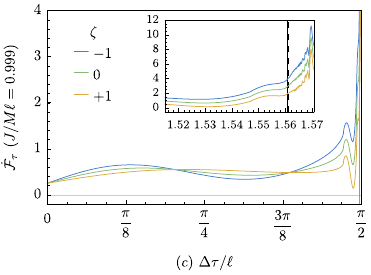}
    \caption{The transition rate $\dot{\mathcal{F}}_\tau$ of a UDW detector freely falling into a rotating BTZ black hole, with $r_0/r_+=100$, $M=10^{-4}$, $E\ell=-5$, and (a) $J/M\ell=0$, (b) $J/M\ell=0.5$, and (c) $J/M\ell=0.999$. We calculated the image sum from $n=-N$ to $n=+N$, where $N=200$ in the exterior of the black hole and $N=1000$ behind the outer horizon. The time for the detector to reach the outer horizon is indicated by the vertical solid line, and the time for the detector to reach the inner horizon is indicated by the vertical dashed line. The right edge of the plot is the time to the singularity for $J=0$. 
    }
    \label{fig:tr-vs-J-Z}
\end{figure}

In Fig. \ref{fig:tr-vs-J-Z}, we show the transition rate $\dot{\mathcal{F}}_\tau$ of the detector as a function of the total detection time $\Delta\tau/\ell$ for the three boundary conditions, $\zeta=-1$, 0, and 1, given black hole angular momentum of (a) $J/M\ell=0$ (static case), (b) $J/M\ell=0.5$, and (c) $J/M\ell=0.999$. In all three plots, we fix the mass of the black hole to be $M=10^{-4}$ and the energy gap of the detector to be $E\ell=-5$.

For the values of $M$, $r_0$, and $J$ considered in Fig.~\ref{fig:tr-vs-J-Z}, the behavior of the transition rate curve changes over different ranges of $\Delta\tau/\ell$, and can be analyzed in three cases: (i) the detector is switched off when it is far away from the black hole, (ii) the detector is switched off when its radial position is comparable to the outer horizon of the black hole, and (iii) the detector is switched off after it has crossed the inner horizon of the black hole.

For short durations of operation, when the detector is far away from the horizon, the transition rate undergoes slow oscillation as dictated by the complex exponential function in the integrand of (\ref{eq:transition_rate}). This oscillation, when computed for the different boundary conditions, will be out of phase for some values of $\zeta$ compared to the others.
For $\Delta\tau/\ell$ corresponding to a detector position near the outer horizon of the black hole, the fluctuations in the transition rate become more dramatic.  Glitches begin to appear in this region, and after their occurrence, the transition rate curves fluctuate with greater frequency and amplitude. 
The increased fluctuation arises from the contributions of the $|n|\geq 1$ terms in the image sum---specifically, when the $n$-th term in the image sum encounters a glitch, its behavior will change from slow and weak oscillation to a more rapid and appreciable oscillation. A detailed discussion on the role of the $|n|\geq 1$ terms is presented in Fig.~\ref{fig:individual_contributions_zeta_0} and Fig.~\ref{fig:individual_contributions_zeta_-1_and_1} of the Appendix.

Outside the inner horizon of the black hole, the transition rate curves are smooth, except for a finite number of glitches. However, if the detector remains on after crossing the inner horizon, its transition rate becomes highly irregular. From the insets in Fig.~\ref{fig:tr-vs-J-Z} (b) and (c), it can be seen that, past the inner horizon, the transition rate curves become ``jagged'' for the rotating case, with many inflection points at extremely close times to one another. As the detector continues and approaches the singularity, we see that the transition rate increases overall, but remains finite at the closest approach. We found, however, that only the transition rate function in the static case diverges at the singularity.

In the region of proper time where $r(\Delta\tau/\ell)$ is near or behind the outer horizon, the effect of varying the boundary conditions is overshadowed by the larger effects noted in the previous paragraphs. Aside from an overall translation in the vertical direction, there is no appreciable qualitative difference between the transition rate curves for $\zeta=-1$, 0, and 1 near or inside of the black hole's outer horizon. For this reason, in the following sections, we fix $\zeta=0$.

Comparing the static case in Fig. \ref{fig:tr-vs-J-Z} (a) with the rotating cases in Fig. \ref{fig:tr-vs-J-Z} (b) and (c), we see that the transition rate curve is similar in all three cases, except for the addition of new behaviors behind the inner horizon when angular momentum is present. The main effect of increasing angular momentum is to push the transition rate curve toward the left (i.e., away from the singularity). We observe that as the angular momentum increases, the detector will encounter the features of the transition rate curve sooner.

\subsection{Transition rate for different black hole masses and angular momenta}

\begin{figure*}[t]
    \centering
    \includegraphics[width=0.49\linewidth]{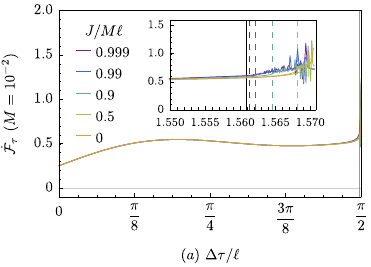}
    \includegraphics[width=0.47\linewidth]{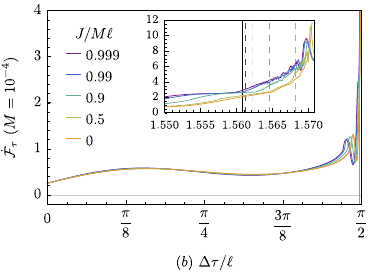}
    \includegraphics[width=0.47\linewidth]{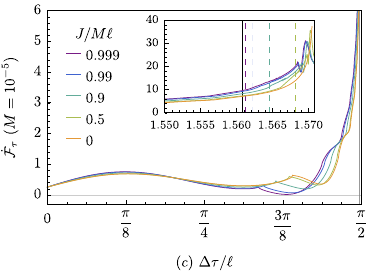}
    \includegraphics[width=0.47\linewidth]{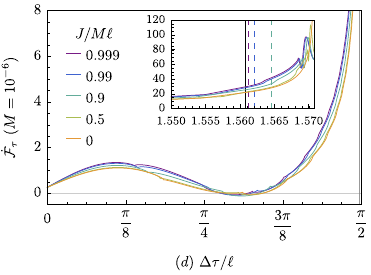}
    \caption{The transition rate $\dot{\mathcal{F}}_\tau$ of a UDW detector freely falling into a rotating BTZ black hole, with $r_0/r_+=100$, $E\ell=-5$, $\zeta=0$, and (a) $M=10^{-2}$, (b) $M=10^{-4}$, (c) $M=10^{-5}$, and (d) $M=10^{-6}$. We calculated the image sum from $n=-N$ to $n=+N$, where $N=200$ in the exterior of the black hole and $N=1000$ behind the outer horizon. The time for the detector to reach the outer horizon is indicated by the vertical solid black line, and the times for the detector to reach the inner horizon are indicated by the vertical dashed lines, where the colors of the dashed lines match the colors of the black hole angular momenta. The right edge of the plot is the time to the singularity for $J=0$.}
    \label{fig:tr-vs-M-J}
\end{figure*}

In Fig. \ref{fig:tr-vs-M-J}, we show the transition rate $\dot{\mathcal{F}}_\tau$ of the detector as a function of the detector's proper time $\Delta\tau/\ell$ for five black hole angular momenta, $J/M\ell=0$, 0.5, 0.9, 0.99, and 0.999, given a black hole mass of (a) $M=10^{-2}$, (b) $M=10^{-4}$, (c) $M=10^{-5}$, and (d) $M=10^{-6}$. In all four plots, we fix the energy gap of the detector to be $E\ell=-5$ and the boundary condition of the field to be $\zeta=0$.

For the values of $M$, $r_0$, and $J$ considered in the previous subsection, we identified three regions of behavior for the transition rate function with respect to the detector operating time. To reiterate, these are (i) the far region, where the detector is in the distant exterior of the black hole, (ii) the near outer horizon region, where the detector crosses or approaches close enough to the outer horizon, and (iii)  the inner horizon region, where the detector has crossed the second horizon of the black hole. Region (i) is characterized by the smooth, slow oscillation of the transition rate function, region (ii) is characterized by larger, faster oscillations interspersed with glitches, and region (iii) is characterized by an extremely high concentration of inflection points, giving the transition rate curve a jagged appearance.

From Fig.~\ref{fig:tr-vs-M-J}, it can be seen that the detector transition rate is highly sensitive to black hole mass. In particular, region (ii) expands outward (to the left) as the black hole mass decreases. Region (i) shrinks as a result, while region (iii) remains confined behind the inner horizon. As region (ii) grows, the detector encounters glitches further and further away from the horizon. Moreover, by adjusting the values of $M$ and $J$, the first type of glitch can be made to appear arbitrarily close to $\Delta\tau/\ell=0$. Thus, for sufficiently small black hole masses, the detection of glitches in the transition rate may serve as an early warning for horizon crossing as proposed in \cite{MariaRosaBTZ}.

In the large mass case ($M=10^{-2}$), region (ii) is virtually nonexistent. The smooth oscillation of the transition rate function continues until the inner horizon, beyond which irregularities begin to appear. In the medium mass case ($M=10^{-4}$), region (ii) has a relatively small extent, stretching from $\Delta\tau/\ell\approx 1.5$ to the inner horizon. Thus, we see slow oscillation for most of the duration of operation, followed by a relatively brief period of more pronounced oscillations, leading into region (iii). We then considered a transitional mass, $M=10^{-5}$, for which the features of the pronounced oscillations in region (ii) develop appreciably and grow leftward. Finally, in the small mass case ($M=10^{-6}$), the detector traverses through region (ii) for most of its operating duration, the first glitch being encountered as early as $\Delta\tau/\ell\approx 5\pi/32$.

The magnitude of the transition rate increases more quickly beyond the horizon when the black hole mass is small. Far away from the horizon, $\dot{\mathcal{F}}_\tau$ oscillates about the value of approximately 0.5, which is true for all four plots in Fig. \ref{fig:tr-vs-M-J}. In the large mass case, the transition rate remains on the order of 1 up to $\Delta\tau/\ell=\pi/2$. However, in the medium mass case, the magnitude of the transition rate grows to roughly 10 in the interval up to $\Delta\tau/\ell=\pi/2$, while in the small mass case, the magnitude grows over 100. We observe again that the transition rate function does not diverge at $\Delta\tau/\ell=\pi/2$ except  when $J/M\ell=0$, the situation in which the freely falling detector reaches the singularity.  We note that the smooth appearance of the transition rate curves behind the inner horizon, for the smaller masses, is purely due to the scaling of the plots. When enlarged sufficiently, one would observe the same jagged shape of the transition rate curve as in the large mass plot.


Finally, we comment on the effect of angular momentum. As was observed in Fig. \ref{fig:tr-vs-J-Z}, increasing the angular momentum of the black hole stretches the transition rate curves away from the singularity. Region (iii) grows outward as the angular momentum pushes the inner horizon toward the outer horizon, and region (ii) also stretches outward to some extent. Consequently, keeping all other parameters constant, the same features of the transition rate curve are encountered by the detector at an earlier proper time when the angular momentum is larger.

\subsection{Transition rate for different detector energy gaps}

\begin{figure}
    \centering
    \includegraphics[width=\linewidth]{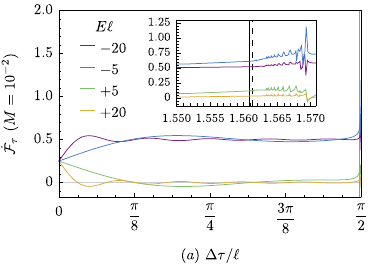}
    \includegraphics[width=\linewidth]{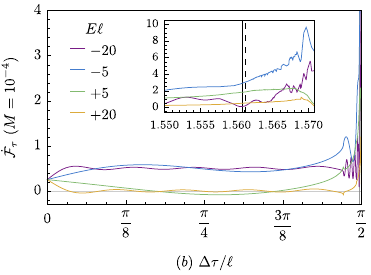}
    \includegraphics[width=\linewidth]{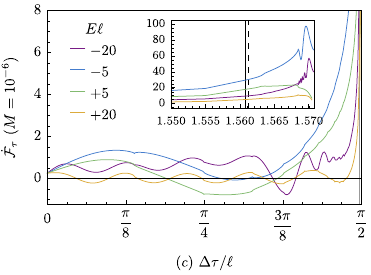}
    \caption{The transition rate $\dot{\mathcal{F}}_\tau$ of a UDW detector freely falling into a rotating BTZ black hole, with $r_0/r_+=100$, $J/M\ell=0.999$, $\zeta=0$, and (a) $M=10^{-2}$, (b) $M=10^{-4}$, and (c) $M=10^{-6}$. We calculated the image sum from $n=-N$ to $n=+N$, where $N=200$ in the exterior of the black hole and $N=1000$ behind the outer horizon. The time for the detector to reach the outer horizon is indicated by the vertical solid line, and the time for the detector to reach the inner horizon is indicated by the vertical dashed line. The right edge of the plot is the time to the singularity for $J=0$.}
    \label{fig:tr-vs-M-E}
\end{figure}

In Fig. \ref{fig:tr-vs-M-E}, we show the transition rate $\dot{\mathcal{F}}_\tau$ of the detector as a function of the detector's proper time $\Delta\tau/\ell$ for four energy gaps, $E\ell=-20$, $-5$, $5$, and $20$, given a black hole mass of (a) $M=10^{-2}$, (b) $M=10^{-4}$, and (c) $M=10^{-6}$. In all three plots, we fix  the angular momentum of the black hole to be $J/M\ell=0.999$ and the boundary condition of the field to be $\zeta=0$.

We find that the transition rate curves for all of the energy gaps considered are consistent with the observations regarding black hole mass variation in the previous subsection. Namely, the smaller the mass of the black hole, the further outward region (ii) grows, which holds for all detector energy gaps, as expected since the energy gap does not affect the location of the glitches. Additionally, the magnitude of the transition rate grows more quickly near the singularity when the black hole mass is small, which is observed for any energy gap considered.

In region (i), which is the weak oscillation portion of $\dot{\mathcal{F}}_\tau$, we find that larger $|E\ell|$ corresponds to faster oscillations. This result is  as expected, as $E$ appears in the argument of the complex exponential function in the integrand of Eq. (\ref{eq:transition_rate}), and thus directly adjusts the oscillation frequency. When the detector is far away from the black hole, the transition rate curves corresponding to $E\ell=+\Omega$ and $E\ell=-\Omega$ are roughly reflections of one another with respect to the horizontal line $\dot{\mathcal{F}}_\tau=0.25$. 

As the detector approaches the black hole, the behavior of the transition rate curves becomes significantly different for different energy gaps, especially when the distance between the detector and the black hole's outer horizon is small.  Although the energy gap does not affect the location of the glitches, its magnitude affects the frequency of the oscillations after each glitch. The insight to be gained is that, when constructing a UDW detector, the value of the energy gap plays a large role in defining what one expects to observe as the detector falls into the black hole.

\begin{figure}
    \centering
    \includegraphics[width=\linewidth]{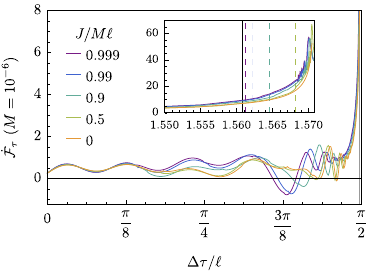}
    \caption{The transition rate $\dot{\mathcal{F}}_\tau$ of a UDW detector freely falling into a rotating BTZ black hole, with $r_0/r_+=100$, $E\ell=-20$, $\zeta=0$, and $M=10^{-6}$. We calculated the image sum from $n=-N$ to $n=+N$, where $N=200$ in the exterior of the black hole and $N=1000$ behind the outer horizon. The time for the detector to reach the outer horizon is indicated by the vertical solid black line, and the times for the detector to reach the inner horizon are indicated by the vertical dashed lines, where the colors of the dashed lines match the colors of the black hole angular momenta. The right edge of the plot is the time to the singularity for $J=0$.}
    \label{fig:TR-vs-M-J-Em20}
\end{figure}

To further connect these results to our observations in the previous subsection, we show, in Fig. \ref{fig:TR-vs-M-J-Em20}, the transition rate $\dot{\mathcal{F}}_\tau$ of the detector as a function of the detector's proper time $\Delta\tau/\ell$ for $E\ell=-20$ (as opposed to $E\ell=-5$ in Fig. \ref{fig:tr-vs-M-J}), given a black hole mass of $M=10^{-6}$. We then compare the transition rates for different black hole angular momenta---specifically, $J/M\ell=0$, 0.5, 0.9, 0.99, and 0.999, as before.

We find that, when the detector is near the outer horizon, the property that increasing angular momentum stretches the transition rate curve to the left is preserved for this new energy gap. However, the behavior in the small mass case ($M=10^{-6}$) is the most complex, due to the far extent of region (ii). Thus, for example, we see a significant discrepancy between the transition rate curves in Fig.~\ref{fig:TR-vs-M-J-Em20} at $\Delta\tau/\ell\approx 3\pi/8$, which arises from the highly nontrivial interaction between the oscillatory properties of the transition rate and the effects of glitches.

\subsection{Glitches in the rotating BTZ problem}

\begin{figure*}[t]
    \centering
    \includegraphics[width=0.49\linewidth]{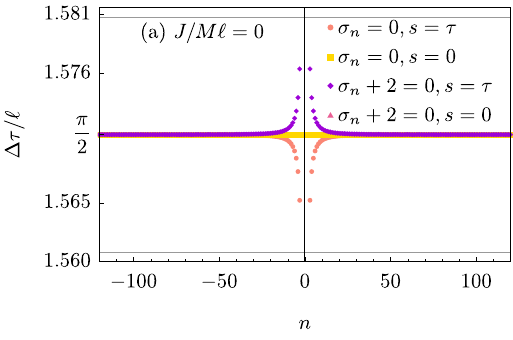}
    \includegraphics[width=0.49\linewidth]{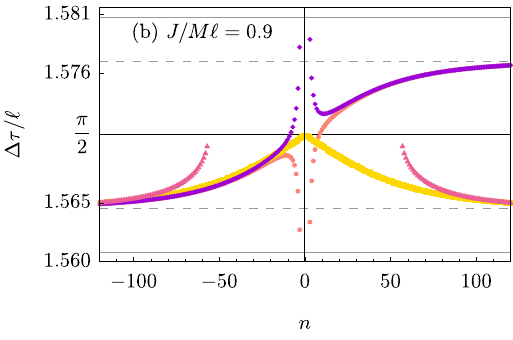}
    \caption{The distribution of glitches in the (a) static case  compared to the (b) rotating case with $J/M\ell=0.9$, where we have fixed $M=10^{-4}$ and $r_0/r_+=100$. We plot detector proper time $\Delta\tau/\ell$ on the vertical axis and term number $n$ on the horizontal axis. The four types of glitches are denoted by different markers.  The circle and square markers correspond to glitches that arise from singularities in the terms independent of the field's boundary conditions in \eqref{eq:transition_rate}, whereas the diamond and triangle markers correspond to glitches that arise from the terms that depend on the boundary conditions. 
    In (a), we have extended the vertical axis past the proper time to the singularity for purposes of illustration. 
    {The lower solid (dashed) horizontal line indicates the time to the outer (inner) horizon. The upper lines are reflections of the time to the horizons with respect to $\Delta\tau/\ell=\pi/2$.} 
    }
    \label{fig:glitches}
\end{figure*}

In Fig. \ref{fig:glitches}, we show the distribution of glitches in the static BTZ case compared to the rotating BTZ case. We set the parameters to be $M=10^{-4}$ and $r_0/r_+=100$, with $J/M\ell=0.9$ for the angular momentum of the black hole in the rotating case. The detector proper time $\Delta\tau/\ell$ is plotted on the vertical axis against $n$ on the horizontal axis, where $n$ is the term number in the image sum. Thus, each point on the plot indicates a glitch at operating time $\Delta\tau/\ell$ arising from the $n$-th term of the summation.

Recall from section II that the four types of glitches arise from setting $s_i^*(\tau)=0$ or $s_i^*(\tau)=\tau$, with the explicit proper times of the glitches given in Eqs (\ref{eq:glitch-explicit-tau1}-\ref{eq:glitch-explicit-tau4}). In the nonrotating case, the glitches are symmetric under $n\rightarrow -n$. In the rotating case, two of the four types of glitches become asymmetric under $n\rightarrow -n$, though they remain symmetric to one another by a reflection through the ``origin,'' which we take to be ($n=0$, $\Delta\tau/\ell=\pi/2)$. The fact that the glitches are not symmetric as $n\rightarrow -n$ in the rotating case is consistent with the fact that $\dot{\mathcal{F}}_\tau$ is in general not symmetric as $n\rightarrow -n$. It is important to note that glitches beyond $\Delta\tau/\ell=\pi/2$ are unphysical for the nonrotating case, as the detector will have already reached the singularity by then, but we include them in Fig. \ref{fig:glitches} to illustrate the mathematical structure of the nondifferentiable points of the transition rate.

In the rotating case, we also observe that three of the four types of glitches are confined behind the inner horizon. Asymptotically, all glitches accumulate at the inner horizon  (or its reflection with respect to $\Delta\tau/\ell=\pi/2$) as $n\rightarrow \pm\infty$. Only one type of glitch (which arises from the first square root for $s_1^*(\tau)=\tau$) is able to escape from the inner horizon for small $|n|$. Thus, in the limit as $J\rightarrow 0$ for the black hole, and the inner horizon goes to the singularity, we are left with only one observable type of glitch. We find that, despite the richer behavior of glitches in the rotating BTZ case, we are able to retrieve the nonrotating case in a continuous limit.

Finally, we reiterate that the proper times at which the detector encounters the glitches depends on the initial position of the detector as well as the mass and angular momentum of the black hole. As mass decreases or angular momentum increases, the glitches will stretch downward in the plots in Fig.~\ref{fig:glitches}, meaning that they will be encountered by the detector at earlier proper times. However, the overall shapes of the curves formed by the four types of glitches will remain the same.

\subsection{Detector transition rate approaching the singularity}

\begin{figure*}[t]
    \centering
    \includegraphics[width=0.49\linewidth]{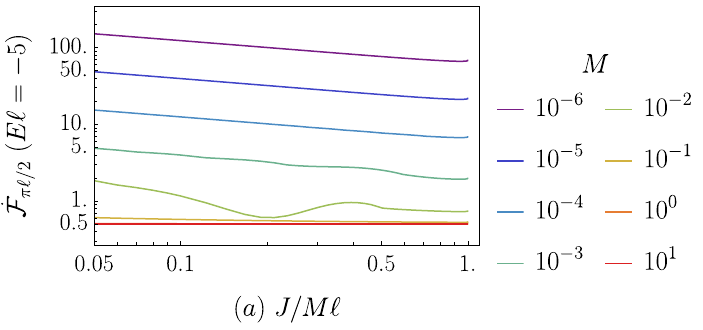}
    \includegraphics[width=0.49\linewidth]{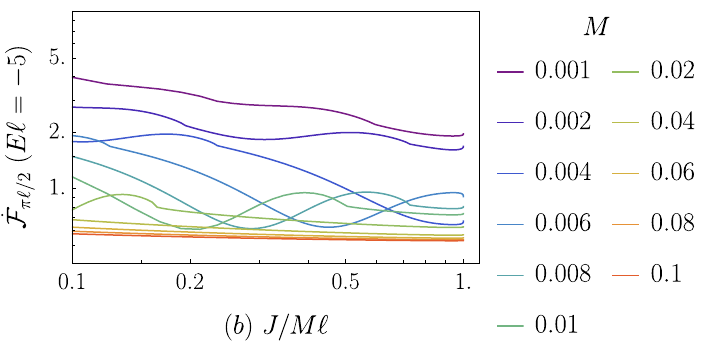}
    \includegraphics[width=0.49\linewidth]{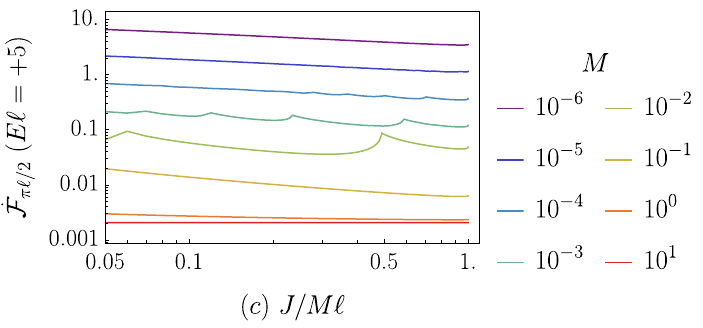}
    \includegraphics[width=0.49\linewidth]{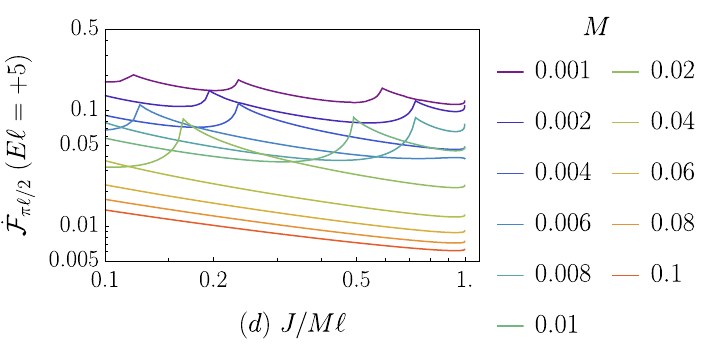}
    \caption{The transition rate   $\dot{\mathcal{F}}_{\pi \ell/2}$ as a function of black hole angular momentum, for different black hole masses. Plots (a) and (b) correspond to a detector energy gap of $E\ell=-5$, while plots (c) and (d) correspond to a gap of $E\ell=+5$. For all plots, $\zeta=0$ and $r_0/r_+=100$. When the black hole mass is large, the transition rate at $\Delta\tau/\ell = \pi/2$ is roughly constant for all angular momenta. When the black hole mass is small, the transition rate at $\Delta\tau/\ell = \pi/2$ decays as an inverse power of $J/M\ell$, as shown by parallel linear traces on the double-log plot (a), (c). However, at intermediate masses, the relationship between $\dot{\mathcal{F}}_{\pi \ell/2}$ and $J/M\ell$ is highly nontrivial, with fluctuations and glitches. Plots (b) and (d) highlight the nontrivial glitch structure within this intermediary band of masses.}
    \label{fig:tr-at-singularity}
\end{figure*}

As observed in previous sections, the transition rate of the detector does not diverge as it approaches the singularity when the black hole is rotating. This absence of a divergence is an important qualitative difference between the static and rotating BTZ scenarios, and arises because the detector's point of closest approach is $r=\frac{r_{-}r_+}{r_0}$ at $\Delta\tau/\ell=\pi/2$.

We conjecture that the 
divergence in the transition rate for vanishing angular momentum  at the time the detector reaches the singularity   is a consequence of the manifold breaking down at the singularity. We find for the rotating black hole that the transition rate is everywhere finite, commensurate with  the manifold being well defined at the singularity.
In Fig. \ref{fig:tr-at-singularity}, we show the transition rate $\dot{\mathcal{F}}_\tau$ of the detector when $\Delta\tau/\ell=\pi/2$ as a function of black hole angular momentum $J/M\ell \in (0,1)$ for black hole mass ranging from $M=10^{-6}$ to $M=10$. We compare the $E\ell=-5$ and $E\ell=+5$ cases.

When the black hole mass is large, the detector's transition rate at $\Delta\tau/\ell=\pi/2$ is essentially independent of the angular momentum $J$. This occurs because the image sum converges more quickly for larger $M$. When $M$ is sufficiently large, the sum is dominated by the $n=0$ term, and thus, in the large mass limit, the transition rate at $\Delta\tau/\ell=\pi/2$ tends to the transition rate in pure AdS$_3$ spacetime. 

When the black hole mass is small, the detector's transition rate at $\Delta\tau/\ell=\pi/2$ decays as an inverse power of $J/M\ell$, as seen from the straight lines in the double-log plot in Fig. \ref{fig:tr-at-singularity} (a), (c). As long as the mass is nonzero, the transition rate remains finite. Numerically we find that the transition rate continues to increase as $M\rightarrow 0$ for all $J$. 

Between the small and large mass extremes is a range of masses for which the relationship between the transition rate at $\Delta\tau/\ell=\pi/2$ and the angular momentum is highly nontrivial. For these masses the transition rate  $\dot{\mathcal{F}}_{\pi \ell/2}$ is a nonmonotonic function  of $J/M\ell$. Moreover, there are glitches in the transition rate, which occur when the geodesic distance under one of the square roots in (\ref{eq:transition_rate}) vanishes at $\tau/\ell=\pi/2$. In Fig. \ref{fig:tr-at-singularity}, we considered the transparent boundary condition $\zeta=0$, and thus the glitches in the plots correspond only to $\sigma_n=0$. However, for the other boundary conditions, one obtains additional glitches when $\sigma_n+2=0$.

The behavior of the transition rate as a function of $J/M\ell$ does not depend qualitatively on the gap of the detector. We obtain the same large, small, and intermediary response structure at closest approach to the singularity for positive detector energy gaps ($E\ell=+5$) and negative ones ($E\ell=-5$). This occurs because the Wightman function depends on $M$ but not on $E$, whereas $E$ controls the phase in (\ref{eq:transition_rate}). Thus, the small and large mass limiting behaviors are independent of $E$. 

Understanding what happens to the rate
at $r=0$ in the rotating case will require placing the detector in non-geodesic motion, a subject for future study.

\section{Conclusion}\label{sec:conclusion}

We numerically calculated the transition rate of an Unruh-DeWitt detector coupled to a massless conformal scalar field as the detector falls freely into a rotating BTZ black hole along a corotating trajectory. We explored the effects of different black hole masses, black hole angular momenta, detector energy gaps, and field boundary conditions at infinity. To calculate the transition rate, we expressed the Wightman function of the BTZ black hole as an image sum of Wightman functions in AdS$_3$, and then computed the transition rate as a sum of the contributions arising from each image. Unlike in the nonrotating case, the terms in the transition rate when rotation is present are not invariant under $n\rightarrow -n$. We found that selecting a different boundary condition for the field at infinity did not change the qualitative behavior of the transition rate curves. However, black hole mass and detector energy gap did significantly impact the transition rate, while black hole angular momentum also played a role. Specifically, the presence of angular momentum caused a stretching of the transition rate curve away from the singularity.

As with the nonrotating BTZ case, we discovered that the detector's transition rate is nondifferentiable at certain discrete values of the detector's proper time. Following the nomenclature in~\cite{MariaRosaBTZ}, we refer to these points of nondifferentiability as glitches. There are four types of glitches in the rotating BTZ case, which are asymmetric for $n\rightarrow -n$. The exotic glitches (i.e., those that were not seen in the static case) live entirely behind the inner horizon and thus vanish as expected as the angular momentum of the black hole goes to zero. The locations of the glitches depend on the black hole mass and angular momentum, and also (though we did not consider it in this study) the initial radial position of the detector.

We observed that, in the presence of rotation, the detector's transition rate at $\Delta\tau/\ell=\pi/2$, at which the detector reaches $r=\frac{r_{-}r_+}{r_0}$ (its point of closest approach to the singularity)
does not diverge. Furthermore, we identified a range of masses for which the transition rate at this point exhibits glitches when plotted as a function of black hole angular momentum. In the small mass limit, the transition rate at $\Delta\tau/\ell=\pi/2$ decays as an inverse power of angular momentum, while in the large mass limit, the transition rate tends to the transition rate in pure AdS$_3$ spacetime.

Our study was motivated first by the observation that the response function of a detector interacting with the Hartle-Hawking(-Israel) state of a massless scalar field and freely falling into a (3+1)-dimensional Schwarzschild black hole exhibits a local extremum at the horizon crossing~\cite{Ng2022}. Following this work with the Schwarzschild black hole was a further study~\cite{MariaRosaBTZ}, wherein a UDW detector was permitted to fall radially into a (2+1)-dimensional nonrotating BTZ black hole. In this latter study, in addition to the discovery of glitches, it was found that more than one local extremum could exist near the horizon under certain parameters. Our results for the (2+1)-dimensional rotating BTZ black hole are commensurate with both of these previous studies and, moreover, provide a more general description of UDW detectors in BTZ spacetime (from which the previous nonrotating BTZ case can be derived). The possibility of using glitches as an early warning for horizon crossing, so long as black hole mass is sufficiently small, remains valid in the presence of rotation. 

One limitation of our study is that we calculated the transition rate of the detector during infall, as opposed to the response function, which is proportional to the transition probability of the detector and thus easier to interpret. This constraint makes the comparison with~\cite{Ng2022} somewhat cumbersome, as the (3+1)-dimensional Schwarzschild study calculated the response function directly. However, we may still observe a number of qualitative similarities between the two sets of results. Another limitation is that we have considered a (2+1)-dimensional spacetime with constant curvature, whereas physical black holes are (3+1)-dimensional and have varying spacetime curvature. Nevertheless, we expect the results of our study to provide useful insight on the 
nature of quantum fields in rotating black hole spacetime backgrounds.

A future avenue of research would be to study the entanglement dynamics of multiple detectors in rotating BTZ spacetime, specifically when one or more of the detectors falls behind the horizon. Additionally, the work on infalling UDW detectors can be extended to other black hole spacetimes where the curvature may not be constant. The role of hidden topologies may also be a topic of interest---for instance, comparing the transition rates of a detector falling into a BTZ black hole versus its geon counterpart. Finally, one could explore the transition probability of a detector falling into a black hole while coupled to other fields---for example, a fermionic field.

\section*{Acknowledgments}

 We thank Jorma Louko for helpful discussions on aspects of this project.
This work was supported in part by the Natural Sciences and Engineering Research Council of Canada.
M.~R. Preciado-Rivas gratefully acknowledges the support from the Mike and Ophelia Lazaridis Graduate Fellowship.

\appendix*

\section{Individual terms}\label{sec:appendix}

\numberwithin{figure}{section}
\setcounter{figure}{0} 
\renewcommand{\thefigure}{A\arabic{figure}}

\begin{figure*}[t]
    \centering
    \includegraphics[width=0.49\linewidth]{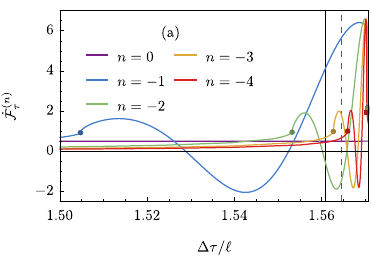}
    \includegraphics[width=0.49\linewidth]{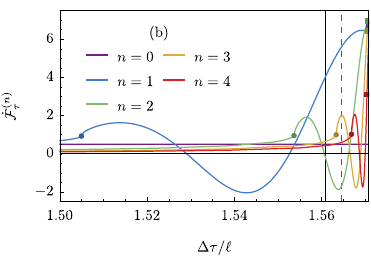}
    \includegraphics[width=0.49\linewidth]{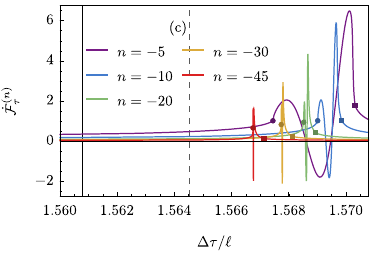}
    \includegraphics[width=0.49\linewidth]{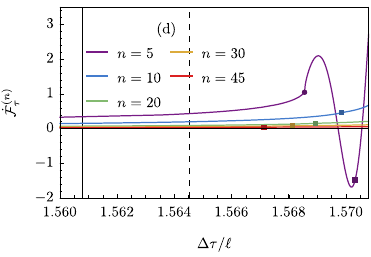}
    \includegraphics[width=0.49\linewidth]{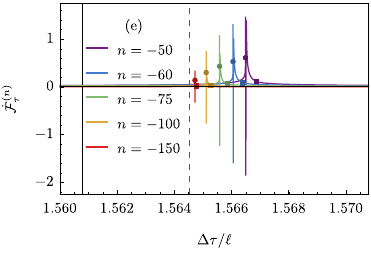}
    \includegraphics[width=0.49\linewidth]{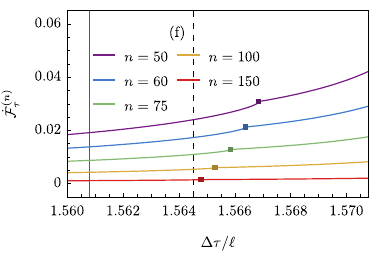}
    \caption{A few individual contributions to the transition rate calculated with $\zeta=0$ 
    and the same parameters as in Fig.~\ref{fig:glitches}. (a), (c), and (e) correspond to terms of negative $n$; meanwhile (b), (d), and (f) correspond to terms of positive $n$. The time at the outer (inner) horizon is indicated with a solid (dashed) vertical line. The glitches are shown with markers following the notation in Fig.~\ref{fig:glitches}. The circle and square markers correspond to the first and second types of glitches, which arise from the terms independent of the boundary condition.}
    \label{fig:individual_contributions_zeta_0}
\end{figure*}

\begin{figure*}[ht]
    \centering
    \includegraphics[width=0.49\linewidth]{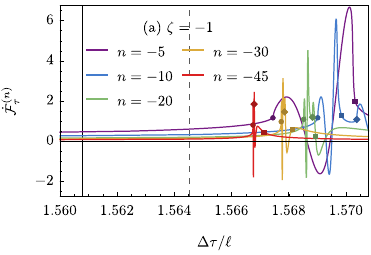}
    \includegraphics[width=0.49\linewidth]{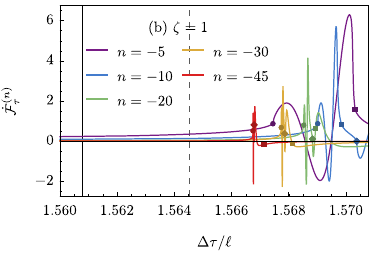}
    \includegraphics[width=0.49\linewidth]{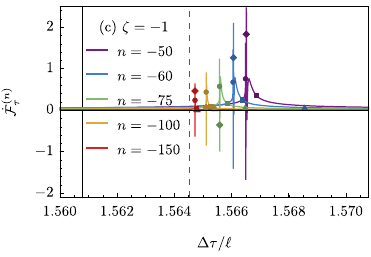}
     \includegraphics[width=0.49\linewidth]{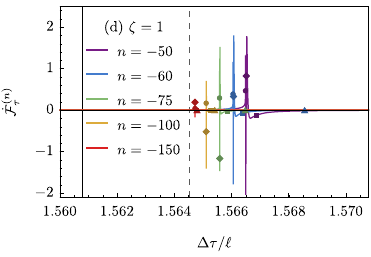}
    \includegraphics[width=0.49\linewidth]{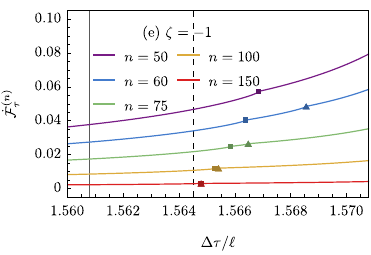}
    \includegraphics[width=0.49\linewidth]{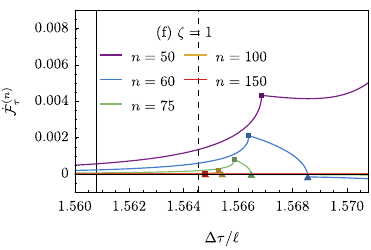}
    \caption{Individual contributions to the transition rate calculated with $\zeta=-1$ or $\zeta=1$ and the same parameters as in Fig.~\ref{fig:glitches}. We show values of $n$ (corresponding to (c), (e) and (f) of Fig.~\ref{fig:individual_contributions_zeta_0}) that differ considerably with respect to the terms calculated with $\zeta=0$. Plots (a), (c), and (e) in this figure depict the $\zeta=-1$ case while (b), (d), and (f) depict the $\zeta=1$ case. The glitches are shown with markers following the notation in Fig.~\ref{fig:glitches}. The diamond and triangle markers correspond to the third and fourth types of glitches, which arise from singularities in the term that depend on the field's boundary conditions.}
\label{fig:individual_contributions_zeta_-1_and_1}
\end{figure*}

In this Appendix, we plot selected individual contributions from terms in the image sum using $M=10^{-4}$ and $J/M\ell=0.9$, shown in Fig.~\ref{fig:individual_contributions_zeta_0} for $\zeta=0$ and in
Fig.~\ref{fig:individual_contributions_zeta_-1_and_1} for $\zeta = \pm 1$.
Following the notation used in Fig.~\ref{fig:glitches}, the circle and square markers indicate glitches that correspond to the first ($\sigma_n(\tau,\tau-s)=0$, $s(\tau)=\tau$) and second type ($\sigma_n(\tau,\tau-s)=0$, $s(\tau)=0$) of glitches, which arise from the terms independent of the boundary conditions. 
The diamond and triangle markers correspond to the third ($\sigma_n(\tau,\tau-s)+2=0$, $s(\tau)=\tau$) and fourth type ($\sigma_n(\tau,\tau-s)+2=0$, $s(\tau)=0$) of glitches, which arise from terms that depend on the boundary conditions. 

As seen in Fig.~\ref{fig:individual_contributions_zeta_0} (a) and (b), the shape of the first few terms is mostly determined by the oscillations following the first type of glitches. On the other hand, there are no oscillations following the second type of glitches, which are located very close to $\Delta\tau/\ell=\pi/2$. For these first terms, the glitches for negative and positive terms occur at times that are relatively close to each other. As a result, the shape of the individual contributions to the transition rate is essentially the same for positive and negative $n$ terms.

As observed in  Fig.~\ref{fig:individual_contributions_zeta_0} (c) and (d), it is for $n=-5$ and $n=5$ that qualitative differences in the contributions become apparent: while the negative $n$ terms attain three local extrema, the positive $n$ terms attain only two or none.
This symmetry loss happens because the glitches of the first type are located in the band between the time to the horizon and the time to $r=\frac{r_{-}r_+}{r_0}$ for negative $n$ but are located past this time  when $n$ is positive. Moreover, these glitches tend to the inner horizon crossing as $n\to-\infty$, which is observed in Fig.~\ref{fig:individual_contributions_zeta_0} (e), but tend to a reflection of the inner horizon crossing about $\pi/2$ for $n\to\infty$.

In Fig.~\ref{fig:individual_contributions_zeta_0} (e) and (f), we observe that the amplitude of the contributions decreases as $|n|$ increases. The terms decay quickly enough to ensure that the transition rate remains finite over the entire domain of the total detection time, including the time at closest approach to the singularity and the horizons. We also observe that the overall shape is maintained but shrunken along the total detection time axis for bigger $|n|$. This behavior produces the jagged region in the transition rate behind the inner horizon crossing.

The role of the third (diamond) and fourth (triangle) types of glitches is illustrated in Fig.~\ref{fig:individual_contributions_zeta_-1_and_1}, where we have chosen to show the plots analogous to Fig.~\ref{fig:individual_contributions_zeta_0} (c),  (e), and (f) because they differ the most when compared to the contributions calculated with $\zeta=0$. 
For small $|n|$ or for positive $n$, glitches of the third type only occur after $\Delta\tau/\ell=\pi/2$. Hence, these contributions do not differ substantially from $\zeta=0$.
However, as the magnitude of $n$ increases, the third type of glitches starts to appear in the contributions for negative $n$ terms, as observed in Fig.~\ref{fig:individual_contributions_zeta_-1_and_1} (a) and (b). As expected, the oscillation behind this type of glitches has an opposite phase for $\zeta=\pm1$.

We observe in Fig.~\ref{fig:individual_contributions_zeta_-1_and_1} (c) and (d) that glitches of both the third and fourth types are present in the terms for negative and large magnitude $n$, but the amplitude of the curves after the fourth type of glitches is negligible compared to the rest of the contribution. Since the glitches of the third type tend to a reflection of the inner horizon crossing with respect to $\pi/2$ as $n\to\infty$, they are not present in the plots of Fig.~\ref{fig:individual_contributions_zeta_-1_and_1} (e) and (f). Although the behavior is qualitatively different for $\zeta=\pm1$, the amplitude of the contributions after the fourth type of glitch is again negligible when compared to other terms.

Overall, we note that the terms that are mostly determined by the first type of glitch, which are the ones with small $|n|$, resemble those that were present in the nonrotating case of \cite{MariaRosaBTZ}. We also note that when more terms are included in the image sum, the difference between the static and the rotating black holes becomes apparent because the exotic types of glitches are present. These new types of glitches, which were absent in the static case, occur close to one another behind the inner horizon crossing and do not pile up at $\Delta\tau/\ell=\pi/2$, which results in transition rate curves that are jagged behind the inner horizon crossing and finite at $\Delta\tau/\ell=\pi/2$.

\newpage

\bibliography{ref}

\end{document}